\documentstyle[psfig]{mn}

\newif\ifAMStwofonts

\newcommand{\lapp}{\mbox{\raisebox{-0.3em}{$\stackrel{\textstyle <}{\sim}$}}}
\newcommand{\gapp}{\mbox{\raisebox{-0.3em}{$\stackrel{\textstyle >}{\sim}$}}}

\title[H{\sc i} gas in rejuvenated radio galaxies]{H{\sc i} gas in the rejuvenated radio galaxy 4C29.30}
\author[Y. Chandola et al.]
  {Yogesh Chandola,$^1$$\thanks{E-mail: chandola@ncra.tifr.res.in (YC), djs@ncra.tifr.res.in (DJS), 
                               Neeraj.Gupta@atnf.csiro.au (NG)  }$ 
    D.J. Saikia$^{1,2}$ and Neeraj Gupta$^2$ \\
$^1$ National Centre for Radio Astrophysics, TIFR, Pune University Campus, Post Bag 3, Pune 411 007, India \\
$^2$ Australia Telescope National Facility, CSIRO, PO Box 76, Epping NSW 1710, Australia}
\date{Accepted.    Received }

\pagerange{\pageref{firstpage}--\pageref{lastpage}}
\pubyear{  }

\begin{document}

\maketitle

\label{firstpage}

\begin{abstract}
We report the results of our observations of H{\sc i} absorption towards the 
central region of the rejuvenated 
radio galaxy 4C29.30 (J0840+2949) with the Giant Metrewave Radio Telescope (GMRT). 
The radio source has diffuse, extended emission with an angular size of $\sim$520 arcsec (639 kpc)
within which a compact edge-brightened double-lobed source with a size of 29 arcsec (36 kpc) is 
embedded. The absorption profile which is seen towards the central component of the inner double
is well resolved and consists of six components; all but one of which appears to be
red-shifted relative to the optical systemic velocity. The neutral hydrogen column density is 
estimated to be $N$(H{\sc i})=4.7$\times$10$^{21}$($T_s$/100)($f_c$/1.0) cm$^{-2}$, where 
$T_s$ and $f_c$ are the spin temperature and covering factor of the background source 
respectively. This detection reinforces a strong correlation between the occurrence of 
H{\sc i} absorption and rejuvenation of radio activity suggested earlier, with the 
possibility that the red-shifted gas is fuelling the recent activity.  
\end{abstract}


\section{Introduction}
One of the interesting and important questions in our understanding of 
active galactic nuclei (AGN) is whether such activity is usually episodic,
and if so, the range of time scales of AGN activity. Besides helping to
constrain models of episodic acitvity, this also has wider
implications in our understanding of AGN feedback in structure formation 
and the evolution of galaxies (e.g. Sijacki et al. 2007, and references
therein; Nesvadba \& Lehnert 2008).  In the presently widely accepted
paradigm, AGN activity is believed to be intimately related to the `feeding' 
of a supermassive black hole whose mass ranges from $\sim$10$^6$ to 
10$^{10}$ M$_\odot$ (e.g. Marconi et al. 2004). Periodic `feeding' of the 
supermassive black hole may lead to different cycles of activity.

\begin{figure*}
\vbox{
   \psfig{file=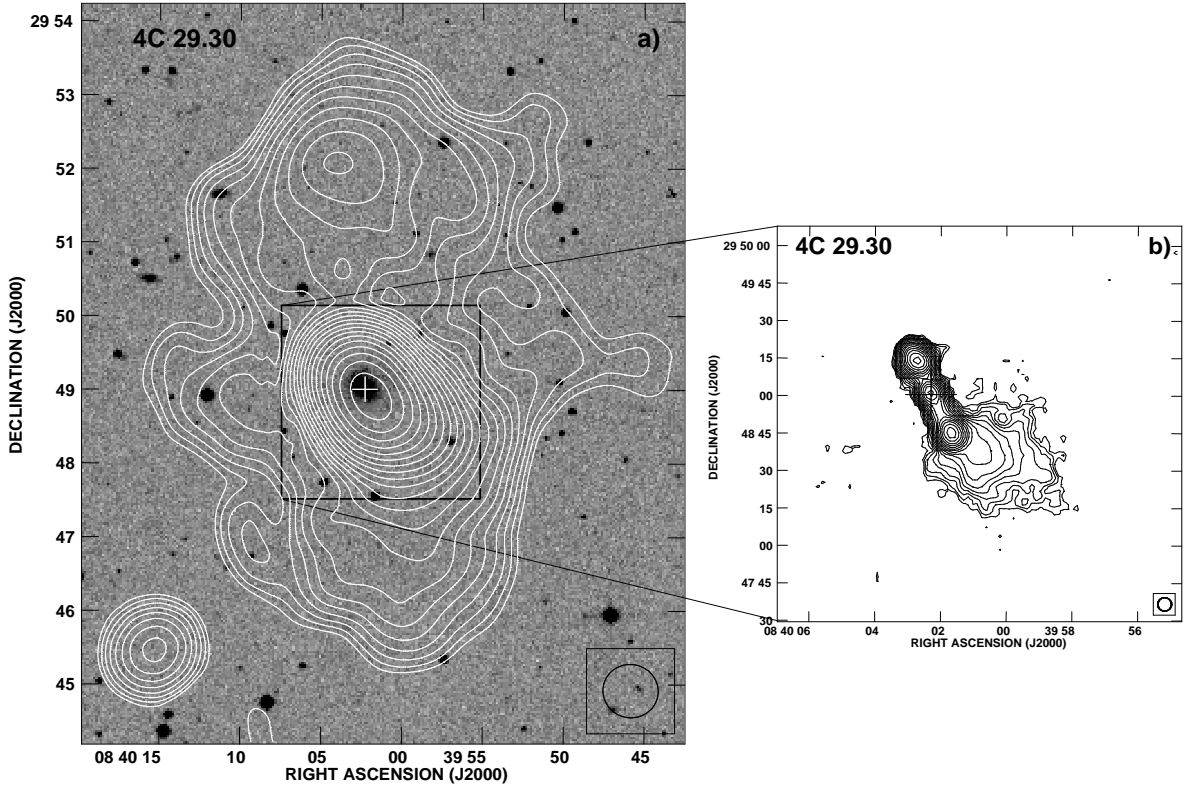,width=6.35in,angle=00}
    }
\caption[]{A collage showing the large- and small-scale structure of the
rejuvenated radio galaxy 4C29.30 (J0840+2949) at 1400 MHz reproduced from Jamrozy et al. 
(2007). The left panel shows the  D-array contour map of the entire source
overlayed on the optical field from the Digital Sky Survey (DSS). The contour levels
are spaced by factors of $\sqrt{2}$ and the first contour is 0.3~mJy~beam$^{-1}$.
The right panel shows the contour map of the central part of the source, 
containing the inner double, from the FIRST (Faint Images of the Radio Sky at Twenty-cm,
Becker, White \& Helfand 1995) survey. The contour
levels are spaced by factors of $\sqrt{2}$, and the first contour is
0.45~mJy~beam$^{-1}$. The size of the beam is indicated by an ellipse in the
bottom right corner of each image. 
         }
\end{figure*}

For radio-loud AGN, an interesting way of probing their history and hence
episodic jet activity is via the structural and spectral information of the lobes of extended 
radio emission.  (e.g.  Burns, Schwendeman \& White 1983; 
Burns, Feigelson \& Schreier 1983; van Breugel \& Fomalont 1984; Leahy, Pooley 
\& Riley 1986; Baum et al. 1990; Clarke, Burns \& Norman 1992; Junkes et al. 1993; 
Roettiger et al. 1994; Schoenmakers et al. 2000; Gizani \& Leahy 2003; Konar et al. 2006).
A very striking example of episodic jet activity is when a new pair of radio lobes
is seen closer to the nucleus before the `old' and more distant radio
lobes have faded (e.g. Subrahmanyan, Saripalli \& Hunstead 1996; Lara et al. 1999).
Such sources have been christened as `double-double' radio galaxies
(DDRGs) by Schoenmakers et al. (2000).
Saikia, Konar \& Kulkarni (2006) reported the discovery of a new DDRG J0041+3224 and
compiled a sample of approximately a dozen such objects from 
the literature, including 3C236 (Schilizzi et al. 2001) and J1247+6723 (Marecki et al. 2003;
Bondi et al. 2004). The inner doubles in these two sources are compact with sizes of 1.7 kpc and
14 pc respectively, and have been classified as a compact steep spectrum (CSS) and 
a Gigahertz peaked spectrum (GPS) source respectively. The median linear size of the 
inner doubles for this sample of DDRGs is $\sim$150 kpc, while the overall median total linear
size is approximately a Mpc. In addition to the classic DDRGs,
evidence of episodic activity may also be seen as diffuse, steep-spectrum emission
from an earlier cycle of activity, in which a young double-lobed radio source may be
embedded. An archetypal example of such a source, namely 4C29.30, was studied in detail
by Jamrozy et al. (2007).

If the nuclear or jet activity is rejuvenated by a fresh supply of gas one might
be able to find evidence of this gas via H{\sc i} absorption towards the radio components
in the central regions of the host galaxy. Saikia, Gupta \& Konar (2007) reported the
detection of H{\sc i} absorption towards the inner double of the DDRG, J1247+6723. 
From the available information in the literature, they also suggested that there could 
be a strong relationship between the detection of H{\sc i} gas and rejuvenation of radio 
activity.  

We present the results of H{\sc i} absorption towards the rejuvenated radio galaxy 4C29.30 
with the Giant Metrewave Radio Telescope (GMRT). 
The total flux density of the inner double at 1287 MHz was estimated by
Jamrozy et al. (2007) to be 390 mJy when observed with an angular resolution of $\sim$2.6 arcsec, 
suggesting it to be a suitable source for these observations. We describe some of the basic 
properties of 4C29.30 (J0840+2949)  in Section 2, the observations
and analyses in Section 3 and present the results and discussions in Section 4. 
The conclusions are summarised in Section 5.

\section{4C29.30 (J0840+2949)}
The radio galaxy 4C29.30 (J0840+2949) is associated with
a bright (R $\rm\sim 15^{m}$) host elliptical galaxy (RA $\rm 08^{h}40^{m}02\fs370$,
DEC $+29\degr49\arcmin02\farcs60$ in J2000 co-ordinates)
at a redshift of 0.064715$\pm$0.000133 as listed in the NASA Extragalactic Database 
from Wegner et al. (2001). The corresponding
luminosity distance is 287 Mpc and 1 arcsec corresponds to 1.228 kpc in a Universe with
H$_\circ$=71 km s$^{-1}$ Mpc$^{-1}$, $\Omega_m$=0.27, $\Omega_\Lambda$=0.73 (Spergel et al. 2003).
The radio images with arcsec resolution show a double-lobed radio source with two
hotspots at the outer edges separated by $\sim$29 arcsec (36 kpc)
and a prominent jet towards the south-west (e.g. van Breugel et al. 1986; Parma et al. 1986
and references therein; Jamrozy et al. 2007).
In addition there is a diffuse blob of emission towards the south-west (SW blob) with a size of
$\sim$40 arcsec (50 kpc) extending beyond the south-western hotspot. These features are all
embedded in diffuse extended emission which has an angular scale of $\sim$520 arcsec (639 kpc),
which has been shown to be due to an earlier cycle of activity (see Fig. 1). 
The inner double has an estimated spectral age of $\lapp$33 Myr, while the diffuse extended emission 
has an estimated spectral age of $\gapp$200 Myr (Jamrozy et al. 2007).


\begin{figure}
\vbox{
    \psfig{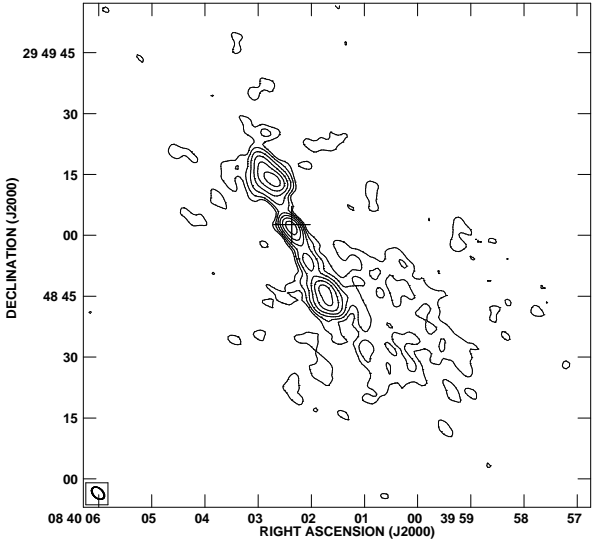}
    }
\caption[]{The GMRT L-band (1332 MHz) image of the inner double from which the 
spectra have been extracted. The rms in the map is 0.4\,mJy\,beam$^{-1}$ and 
the contour levels are 1.5$\times$(-1, 1, 2, 4, 8, 16, 32)\,mJy\,beam$^{-1}$. The 
cross marks the position of the optical galaxy.
          }
\end{figure}

The radio luminosity of the inner double at 1400 MHz is
5.5$\times$10$^{24}$ W Hz$^{-1}$,  which is significantly below the dividing line of the Fanaroff-Riley
classes, while that of the entire source is 7.4$\times$10$^{24}$ W Hz$^{-1}$.  In some of the DDRGs,
the luminosity of the inner double is in the FRI category although its structure resembles that of
FRII radio sources (cf. Saikia et al. 2006).  van Breugel et al. (1986) have shown the presence of 
optical line-emitting gas adjacent to the radio jet along a
position angle (PA) of $\sim$20$^\circ$ and evidence of the radio jets
interacting with dense extranuclear gas. Neutral hydrogen gas has been detected in absorption
with an angular resolution of 3.6 arcmin using the Arecibo telescope by Mirabel (1990), although
the resolution is somewhat coarse to identify the absorber against any of the main radio components.

The host galaxy of 4C29.30 appears to have
merged with a gas-rich galaxy, shows presence of shells and dust
(Gonzalez-Serrano, Carballo \& Perez-Fournon 1993) and is associated with an IRAS source
F08369+2959 (Keel et al. 2005). It is possible that this interaction may have been 
responsible for the rejuvenated activity seen in this source. At X-ray wavelengths {\it Chandra} 
detects emission from the hot spots in the southwestern radio lobe and also in the counterlobe
(Gambill et al. 2003; Sambruna et al. 2004). Both hotspots have also been detected in
observations with the Hubble Space Telescope (Sambruna et al. 2004).
Jamrozy et al. (2007) suggest that the X-ray emission from the hotspots consists of a 
mixture of non-thermal and thermal components, the latter being possibly due to gas 
which is shock heated by the jets from the host galaxy.


\begin{figure}
\vbox{
    \psfig{file=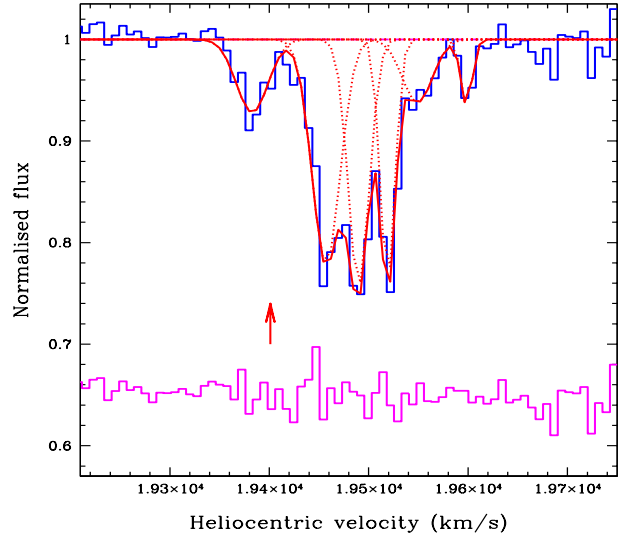,height=2.95in,width=3.35in,angle=0}
    }
\caption[]{The H{\sc i} absorption spectrum (histogram) towards the central component of the radio 
galaxy 4C29.30 (J0840+2949). The Gaussian components fitted to the absorption profile and the sum 
of these components i.e. the fit are plotted as dotted and continuous lines respectively. The
histogram at the bottom shows the residuals shifted upwards. The arrow marks the systemic velocity
as measured from the optical emission lines.
}
\end{figure}



\begin{figure}
\vbox{
    \psfig{file=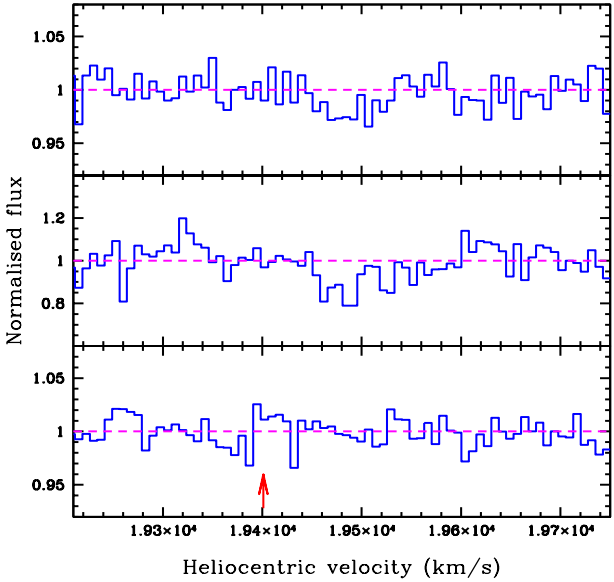,height=3.2in,width=3.35in,angle=0}
    }
\caption[]{The H{\sc i} absorption spectra (histogram) towards the northern hotspot (upper panel), 
southern knot (middle panel) and the southern hotspot (lower panel) of the inner double. 
There is no evidence of any significant absorption detected towards these radio components.
The arrow marks the systemic velocity as measured from the optical emission lines.
}
\end{figure}


\section{Observations and analyses} 
We observed 4C29.30 (J0840+2949)  with the GMRT to search for
associated 21-cm absorption towards the nuclear region and possibly from the hotspots 
of the inner double. The source was observed on 2009 May 05
with a bandwidth of 4 MHz for $\sim$6 hours including calibration overheads.
Approximately 4 hours were spent observing the source.
The 4-MHz bandwidth consisted of 128 spectral channels, giving a spectral  
resolution of $\sim$7 km s$^{-1}$.
The local oscillator chain and FX correlator
system were tuned to centre the baseband bandwidth at 1334 MHz,
the red-shifted 21-cm frequency corresponding to $z_{em}$=0.0647.  We observed the standard
flux density calibrators 3C147 and 3C286 every 3 hours to correct for the variations
in amplitudes and bandpass.  The compact radio source J0741+312 was observed approximately
every 45 minutes for phase calibration of the array.

The data were reduced in the standard way using Astronomical Image Processing System ({\tt AIPS})
package.  After the initial flagging or editing of bad data and calibration, source and
calibrator data were examined for baselines and timestamps affected by Radio Frequency
Interference (RFI).  These data were excluded from further analysis.  A continuum image
of the source was made using calibrated data averaged over the line-free channels.
Due to our interest in absorption towards the compact central source, the imaging
was done without any {\it uv} cut-off or tapering in the visibility plane. This provided us
with the highest possible resolution.
This image was then self-calibrated until a satisfactory map was obtained. 
The self-calibration complex gains determined from this were applied to all the 128
frequency channels and continuum emission was subtracted from this visibility data cube using
the same map. Spectra at peak position of the central source, the hotspots of the inner
double and the knot in the southern jet were extracted from the cube.

\section{Results and discussion}

Our GMRT image of the source (Fig. 2) with an angular resolution of
3.60$\times$2.35 arcsec$^2$ along a position angle of 45.7$^\circ$ and an rms
noise of 0.4 mJy beam$^{-1}$ detects the inner double with a peak 
brightness of 78.9  mJy beam$^{-1}$ for the central component and 74.6 and
71.5 mJy beam$^{-1}$ for the northern and southern hotspots of the inner double. 
The peak brightness of the knot in the jet towards the south is 15.8 mJy beam$^{-1}$.

The H{\sc i} absorption spectrum towards the central component is 
presented in Fig. 3.  H{\sc i} absorption has been detected clearly
towards the core of this rejuvenated radio galaxy. The rms noise in the spectrum 
is $\sim$1 mJy beam$^{-1}$ channel$^{-1}$.  The absorption profile consists of a 
number of components all but one of which appear red-shifted relative to the 
optical systemic velocity.  The H{\sc i} column density, $N$(H{\sc i}), 
for the different spectral components has been estimated using the relation
\begin{equation}
$N$({\rm H{\sc i}})=1.93\times10^{18}\frac{{T}_{s}~\tau_p~\Delta v}{f_c}~ {\rm cm^{-2}},
\label{eq1}
\end{equation}
where $T_s$, $\tau_p$, $\Delta$$v$ and $f_c$ are the spin temperature, peak optical depth,
the full width at half maximum (FWHM) of the Gaussian line profile, 
and the fraction of background emission covered by the absorber respectively.
The estimates have been made assuming $T_s$=100 K and $f_c$=1.0.  The value of 
$T_s$ could be significantly greater than 100 K. For example, 
for the warm neutral medium seen in the Galaxy $T_s$ ranges from 5000$-$8000 K 
(Kulkarni \& Heiles 1988). Such high spin temperatures are also expected to arise 
in the proximity of an active nucleus (Bahcall \& Ekers 1969; Holt et al. 2006). 
The best fit to the 
spectrum with six Gaussian components is shown in Fig. 2 and the fit parameters
are summarised in Table 1. The sixth component requires confirmation from more
sensitive observations. The Table lists the optical heliocentric velocity, 
v$_{\rm{hel}}$, FWHM of the Gaussian profile, the fractional absorption and the
H{\sc i} column density. The numbers within brackets are the errors on the quoted 
values. The  H{\sc i} column density
integrated over the entire absorption profile is 4.7$\times$10$^{21}$ cm$^{-2}$. 

The H{\sc i} absorption spectra towards the hotspots and the `southern knot' 
are shown in Fig. 4.  No significant absorption has been 
detected towards the northern and sourthern hotspots of the inner double or the 
`southern knot' indicating a 3-$\sigma$ upper limits to H{\sc i} of 
$N$(H{\sc i})=9.5$\times$10$^{20}$ cm$^{-2}$, $N$(H{\sc i})=8.3$\times$10$^{20}$ cm$^{-2}$
and $N$(H{\sc i})=5.5$\times$10$^{21}$ cm$^{-2}$ respectively, assuming $T_s$=100 K, $f_c$=1.0
and $\Delta$$v$=100 km s$^{-1}$. This indicates that the size of the absorber
is less than $\sim$35 kpc. However,
the spectra of the `southern knot' and the northern hotspot show very marginal signs 
of absorption at $\sim$1.948$\times$10$^4$ km s$^{-1}$ which needs to be confirmed 
from more sensitive observations. 

\subsection{H{\sc i} gas and rejuvenation of AGN activity}
Saikia et al. (2007) explored any possible relationship between rejuvenation of radio 
or jet activity and the occurrence of H{\sc i}. Unfortunately the number of sources
is still small because most of the rejuvenated radio sources have weak radio emission 
in the central or nuclear
region. In the list of DDRGs compiled by Saikia et al. (2006) the two exceptions are 
3C236 and J1247+6723, with the flux density within a few kpc of the nuclear region 
being $\gapp$100 mJy. The DDRG 3C236 which is a giant radio galaxy with a projected
linear size of $\sim$4250 kpc shows evidence of star formation and 
H{\sc i} absorption against a lobe of the inner radio source (Conway \& Schilizzi 2000; 
Schilizzi et al. 2001; O'Dea et al. 2001). Observations with milliarcsec resolution
are required to determine the location of the absorbing clouds in the case of J1247+6723
reported by Saikia et al. (2007). An interesting case is 3C293, which could be classified
as a misaligned DDRG, and exhibits absorption features both blue- and red-shifted relative
to the the systemic velocity
(Beswick, Pedlar \& Holloway 2002; Beswick et al. 2004), and fast outflowing gas
blue-shifted by upto $\sim$1000 km s$^{-1}$ (Morganti et al. 2003; Emonts et al. 2005).
The case for 3C258 (J1124+1919) as a rejuvenated galaxy is less clear. Although H{\sc i}
absorbing gas was reported by Gupta et al. (2006) towards the compact central source
(e.g. Sanghera et al. 1995), we have not been able to confirm 
from GMRT observations (Saikia et al., in preparation) the weak extended emission reported 
by Strom et al. (1990). Another celebrated case is Centaurus A, where the inner double
is due to more recent activity while the extended diffuse emission is due to earlier
cycles of nuclear activity (Burns et al. 1983; Junkes et al. 1993; Ilana Feain, 
private communication), and H{\sc i} absorption is seen towards the central region
(e.g. Sarma, Troland \& Rupen 2002; Morganti et al. 2008).

While the number of sources is still small, the detection of absorbing H{\sc i}
gas in the rejuvenated galaxies appears to be more frequent than for CSS and GPS objects
(Vermeulen et al. 2003; Gupta et al. 2006), and the reported observations of 4C29.30
reinforces this trend. Although the sample size needs to be clearly increased,
this trend is unlikely to be due to different source strengths. Considering the GPS 
objects listed by Gupta et al. (2006), which has the highest H{\sc i} detection rate 
of $\sim$45 per cent, the median total flux density of the sources at $\sim$1400 MHz 
is $\sim$2 Jy. Amongst the rejuvenated galaxies discussed here, the flux densities 
at $\sim$1400 MHz of the entire central source of 3C236, 3C293 and Cen A are comparable 
to those of the GPS objects, while those of 4C29.30 and J1247+6723 are weaker than 
the GPS objects listed by Gupta et al. (2006). There should not be any bias 
because of source strength. Considering the H{\sc i} column densities (Gupta et al. 2006),
the median value for GPS sources is $\sim$3$\times$10$^{20}$ cm$^{-2}$. The rejuvenated
radio galaxies discussed here have column densities in the range of $\sim$8$-$50$\times$10$^{20}$ 
cm$^{-2}$.


\begin{table}
\caption{Multiple Gaussian fit to the H{\sc i} absorption spectrum towards the 
central component (Fig. 3).}
\begin{center}
\begin{tabular}{|c|l|l|c|c|}
\hline
Id. &   v$_{\rm{hel}}$ &  FWHM      & Frac. abs. & $N$(H{\sc i}) \\
no. &               &               & & 10$^{20}$  \\  
    &   km s$^{-1}$ &   km s$^{-1}$ & & cm$^{-2}$ \\
\hline
1  &  19383(2)  &   37(5)  &    0.072(0.008)   &    5.4(1.4)  \\
2  &  19457(2)  &   33(3)  &    0.222(0.009)   &   16.1(2.0)  \\
3  &  19490(1)  &   25(3)  &    0.243(0.013)   &   13.4(2.1)  \\
4  &  19519(1)  &   18(2)  &    0.238(0.018)   &    9.3(1.6)  \\
5  &  19549(4)  &   36(11) &    0.062(0.009)   &    4.4(2.0)  \\
6  &  19599(1)  &   14(4)  &    0.065(0.013)   &    1.8(1.0)  \\
\hline
\end{tabular}
\end{center}
\label{gauss}
\end{table}


\section{Summary}
We have reported the results of 21-cm absorption towards the central component
of the rejuvenated radio galaxy 4C29.30 (J0840+2949) using 
the GMRT. The absorption profile towards the central component is best fitted by six
components, all but one of which are red-shifted relative to the systemic velocity
of the parent galaxy. The total absorbing neutral hydrogen column density of the gas 
is estimated to be
$N$(H{\sc i})=4.7$\times$10$^{21}$($T_s$/100)($f_c$/1.0)$^{-1}$ cm$^{-2}$. 
The largely red-shifted gas may be responsible for fuelling the black hole causing
the renewed activity in this source. This may have been triggered by 
the interaction of the host galaxy of 4C29.30 
with a gas-rich galaxy which appears to have merged with it
(Gonzalez-Serrano et al. 1993). 

No significant absorbing gas is detected towards either the northern and southern 
hotspots of the inner double, or the knot in the jet towards the south, the 
3-$\sigma$ upper limits being  
$N$(H{\sc i})=9.5$\times$10$^{20}$ cm$^{-2}$, $N$(H{\sc i})=8.3$\times$10$^{20}$ cm$^{-2}$
and $N$(H{\sc i})=5.5$\times$10$^{21}$ cm$^{-2}$ respectively, assuming $T_s$=100 K, $f_c$=1.0
and $\Delta$$v$=100 km s$^{-1}$. 
The possible very weak absorption towards the `southern knot' and the northern
hotspot needs confirmation from more sensitive observations.
The detection of H{\sc i} gas in this rejuvenated radio galaxy reinforces
a possible close relationship between the occurrence of associated H{\sc i} absorption
due to gas clouds and evidence of renewed activity in powerful 
radio galaxies. Some of these absorbing gas clouds may be responsible for 
`feeding' the supermassive black hole with a fresh supply of gas to rejuvenate 
the nuclear radio activity. 

\section*{Acknowledgments} 
YC thanks Aritra Basu, Vishal Gajjar, Nirupam Roy and Sandeep Sirothia for 
their suggestions during data analysis. We thank the referee for a few helpful 
suggestions, and the GMRT staff for help with the observations.  The GMRT is 
a national facility operated by NCRA 
of the Tata Institute of Fundamental Research.
This research has made use of the NASA/IPAC extragalactic database (NED)
which is operated by the Jet Propulsion Laboratory, Caltech, under contract
with the National Aeronautics and Space Administration. We thank numerous 
contributors to the GNU/Linux group. 

{}


\begin{thebibliography}{}

\bibitem[]{}  Bahcall J.N., Ekers R.D., 1969, ApJ, 157, 1055
\bibitem[]{}  Baum S.A., O'Dea C.P., de Bruyn A.G., Murphy D.W., 1990, A\&A, 232, 19
\bibitem[]{}  Becker R.H., White R.L., Helfand D.J., 1995, ApJ, 450, 559
\bibitem[]{}  Beswick R.J., Pedlar A., Holloway A.J., 2002, MNRAS, 329, 620
\bibitem[]{}  Beswick R.J., Peck A.B., Taylor G.B., Giovannini G., 2004, MNRAS, 352, 49
\bibitem[]{}  Bondi M., March\~a, M.J.M., Polatidis A., Dallacasa D., Stanghellini C., Ant\'on S., 2004, 
              MNRAS, 352, 112
\bibitem[]{}  Burns J.O., Feigelson E.D., Schreier E.J., 1983, ApJ, 273, 128
\bibitem[]{}  Burns J.O., Schwendeman E., White R.A., 1983, ApJ, 271, 575
\bibitem[]{}  Clarke D.A., Burns J.O., Norman M.L., 1992, ApJ, 395, 444
\bibitem[]{}  Conway J.E., Schilizzi R.T., 2000, in Conway J.E., Polatidis A.G., Booth R.S., Pihlström Y.M., eds, 
              EVN Symp. 2000, Onsala Space Observatory, p. 123
\bibitem[]{}  Emonts B.H.C., Morganti R., Tadhunter C.N., Oosterloo T.A., Holt J., van der Hulst J.M., 
              2005, MNRAS, 362, 931
\bibitem[]{}  Gambill J.K., Sambruna R.M., Chartas G., Cheung C.C., Maraschi L., Tavecchio F., Urry C.M.; 
              Pesce J.E. 2003, A\&A, 401, 505
\bibitem[]{}  Gizani N.A.B., Leahy J.P., 2003, MNRAS, 342, 399
\bibitem[]{}  Gonzalez-Serrano J.I., Carballo R., Perez-Fournon I., 1993, AJ, 105, 1710	
\bibitem[]{}  Gupta N., Salter C.J., Saikia D.J., Ghosh T., Jeyakumar S., 2006, MNRAS, 373, 972  
\bibitem[]{}  Holt J., Tadhunter C., Morganti R., Bellamy M., Gonz\'alez Delgado R.M., Tzioumis A., 
              Inskip K.J., 2006, MNRAS, 370, 1633
\bibitem[]{}  Jamrozy M., Konar C., Saikia D.J., Stawarz {\L}., Mack K.-H., Siemiginowska A., 2007, MNRAS, 378, 581
\bibitem[]{}  Junkes N., Haynes R.F., Harnett J.I., Jauncey D.L., 1993, A\&A, 269, 29
\bibitem[]{}  Keel W.C., Irby B.K., May A., Miley G.K., Golombek D., de Grijp M.H.K., Gallimore J.F., 
              2005, ApJS, 158, 139
\bibitem[]{}  Konar C., Saikia D.J., Jamrozy M., Machalski J., 2006, MNRAS, 372, 693 (astro-ph/0607660)
\bibitem[]{}  Kulkarni S.R., Heiles C., 1988, in Galactic and Extragalactic Radio Astronomy,
              ed. G. Verschuur \& K. Kellerman (Heidelberg: Springer), 95
\bibitem[]{}  Lara L., M\'arquez I., Cotton W.D., Feretti L., Giovannini G., Marcaide J.M., Venturi T.,
              1999, A\&A, 348, 699
\bibitem[]{}  Leahy J.P., Pooley G.G., Riley J.M., 1986, MNRAS, 222, 753                
\bibitem[]{}  Marconi A., Risaliti G., Gilli R., Hunt L.K., Maiolino R., Salvati M., 2004, MNRAS, 351, 169
\bibitem[]{}  Marecki A., Barthel P.D., Polatidis A., Owsianik I., 2003, PASA, 20, 16 
\bibitem[]{}  Mirabel I.F., 1990, ApJ, 352, L37
\bibitem[]{}  Morganti R., Oosterloo T.A., Emonts B.H.C., van der Hulst J.M., Tadhunter C.N., 2003, ApJ, 593, 69
\bibitem[]{}  Morganti R., Oosterloo T., Struve C., Saripalli L., 2008, A\&A, 485, L5
\bibitem[]{}  Nesvadba N.P.H., Lehnert M.D., 2008, in Charbonnel C., Combes F., Samadi R., eds. 
	      SF2A-2008: Proceedings of the Annual meeting of the French Society of Astronomy and Astrophysics 
	      (http://proc.sf2a.asso.fr), p. 377
\bibitem[]{}  O'Dea C.P., Koekemoer A.M., Baum S.A., Sparks W.B., Martel A.R., Allen M.G., Macchetto F.D., 
              Miley G.K., 2001, AJ, 121, 1915
\bibitem[]{}  Parma P., de Ruiter H.R., Fanti C., Fanti R., 1986, A\&AS, 64, 135 
\bibitem[]{}  Roettiger K., Burns J.O., Clarke D.A., Christiansen W.A., 1994, ApJ, 421, 23L
\bibitem[]{}  Saikia D.J., Konar C., Kulkarni V.K., 2006, MNRAS, 366, 1391
\bibitem[]{}  Saikia D.J., Gupta N., Konar C., 2007, MNRAS, 375, L31
\bibitem[]{}  Sambruna R.M., Gambill J.K., Maraschi L., Tavecchio F., Cerutti R., Cheung C.C., 
              Urry C.M., Chartas G., 2004, ApJ, 608, 698
\bibitem[]{}  Sanghera H.S., Saikia D.J., Luedke E., Spencer R.E., Foulsham P.A., Akujor C.E., Tzioumis A.K., 1995,
              A\&A, 295, 629 
\bibitem[]{}  Sarma A.P., Troland T.H., Rupen M.P., 2002, ApJ, 564, 696
\bibitem[]{}  Schilizzi R.T. et al., 2001, A\&A, 368, 398
\bibitem[]{}  Schoenmakers A.P., de Bruyn A.G., R\"{o}ttgering H.J.A., van der Laan H., Kaiser C.R., 
              2000, MNRAS, 315, 371
\bibitem[]{}  Sijacki D., Springel V., di Matteo T., Hernquist, L. 2007, MNRAS, 380, 877
\bibitem[]{}  Spergel D.N. et al., 2003, ApJS, 148, 175
\bibitem[]{}  Strom R.G., Riley J.M., Spinrad H., van Breugel W.J.M., Djorgovski S., Liebert J., McCarthy P.J., 
              1990, A\&A, 227, 19
\bibitem[]{}  Subrahmanyan R., Saripalli L., Hunstead R.W., 1996, MNRAS, 279, 257
\bibitem[]{}  van Breugel W., Fomalont E.B., 1984, ApJ, 282, 55L
\bibitem[]{}  van Breugel W.J.M., Heckman T.M., Miley G.K., Filippenko A.V. 1986, ApJ, 311, 58
\bibitem[]{}  Vermeulen R.C. et al., 2003, A\&A, 404, 861
\bibitem[]{}  Wegner G., et al. 2001, AJ, 122, 2893 

\end{thebibliography}
\end{document}